\documentclass[preprint2]{aastex}

\usepackage{amsmath,amssymb,graphicx,natbib}

\newcommand{\kepler}{\textit{Kepler}}
\newcommand{\hj}{hot Jupiter}
\newcommand{\wj}{warm Jupiter}
\newcommand{\hn}{hot Neptune}

\newcommand{\rj}{$R_{\text{Jup}}$}
\newcommand{\re}{$R_\oplus$}
\newcommand{\mj}{$M_{\text{Jup}}$}
\newcommand{\me}{$M_\oplus$}

\shorttitle{Hot Jupiters are Lonely}
\shortauthors{Steffen \textit{et al.}}

\author{
Jason H. Steffen\altaffilmark{1},
Darin Ragozzine\altaffilmark{2},
Daniel C. Fabrycky\altaffilmark{3},
Joshua A. Carter\altaffilmark{2},
Eric B. Ford\altaffilmark{4},
Matthew J. Holman\altaffilmark{2},
Jason F. Rowe\altaffilmark{5},
William F. Welsh\altaffilmark{6},
William J. Borucki\altaffilmark{5},
Alan P. Boss\altaffilmark{7},
David R. Ciardi\altaffilmark{8}, and 
Samuel N. Quinn\altaffilmark{2}
}
\altaffiltext{1}{Fermilab Center for Particle Astrophysics, P.O. Box 500, Batavia, IL 60510}
\altaffiltext{2}{Harvard-Smithsonian Center for Astrophysics, 60 Garden St., Cambridge, MA 02138, USA}
\altaffiltext{3}{Department of Astronomy and Astrophysics, University of California, Santa Cruz, Santa Cruz, CA 95064, USA}
\altaffiltext{4}{Astronomy Department, University of Florida, 211 Bryant Space Sciences Center, Gainesville, FL 32111, USA}
\altaffiltext{5}{NASA Ames Research Center, Moffett Field, CA, 94035, USA}
\altaffiltext{6}{Astronomy Department, San Diego State University, San Diego, CA 92182, USA}
\altaffiltext{7}{Department of Terrestrial Magnetism Carnegie Institution for Science, 5241 Broad Branch Road, N.W., Washington, D.C. 20015 USA}
\altaffiltext{8}{NASA Exoplanet Science Institute/California Institute of Technology, Pasadena, CA 91125 USA}

\begin{document}

\bibliographystyle{plainnat}

\title{Kepler constraints on planets near hot Jupiters}

\begin{abstract}
We present the results of a search for planetary companions orbiting near \hj\ planet candidates (Jupiter-size candidates with orbital periods near 3 days) identified in the \kepler\ data through its sixth quarter of science operations.  Special emphasis is given to companions between the 2:1 interior and exterior mean-motion resonances.  A photometric transit search excludes companions with sizes ranging from roughly 2/3 to 5 times the size of the Earth, depending upon the noise properties of the target star.  A search for dynamically induced deviations from a constant period (transit timing variations or TTVs) also shows no significant signals.  In contrast, comparison studies of warm Jupiters (with slightly larger orbits) and hot Neptune-size candidates do exhibit signatures of additional companions with these same tests.  These differences between \hj s and other planetary systems denote a distinctly different formation or dynamical history.
\end{abstract}


\maketitle

Considerable observational evidence indicates that \hj\ planets may constitute a relatively small population with a nonstandard dynamical history; the origins of that population remain unclear.  The ``pile-up'' of Jupiter mass planets with orbital periods between 1 to 5 days has long been noted \citep{Cumming:1999,Udry:2007,Latham:2011}.  The number of \hj s decline rapidly as masses exceed 2\mj\ \citep{Zucker:2002}, and planets with much smaller masses or sizes do not appear to have a similar pile-up.  
Here we study a sample of candidate \hj\ systems from the \kepler\ catalog presented in \citep{Borucki:2011} (hereafter B11).  At the same time, comparison samples of warm Jupiters with slightly longer orbital periods and smaller, ``\hn'' systems are chosen and studied in similar fashion (c.f., Figures \ref{sizecut} and \ref{fullsample}) and are used to demonstrate the differences between these and the \hj\ candidates.

Two broad classes of models seek to explain the origin of the \hj\ population.  One model invokes dynamical perturbations that induce a large eccentricity in the orbit of the Jupiter \citep{Rasio:1996,Weidenschilling:1996,Holman:1997}, after which the semi-major axis and eccentricity are damped by tidal dissipation \citep{Wu:2003,Fabrycky:2007,Nagasawa:2008}.  In the second method, a Jovian planet migrates through a gas disk \citep{Goldreich:1980,Ward:1997}, stopping close to the host star either by a magnetospheric cavity clearing the disk material, Roche lobe overflow \citep{Trilling:1998}, or by the planet raising tides on the star which then injects energy into the planetary orbit---in a fashion similar to the Earth-moon system---preventing its further decay \citep{Lin:1996}.

For the second method, regardless of the stopping mechanism, the time that migration stops will be different for the various planets within a single system as each planet's location and mass is unique.  Consequently, disk-embedded low mass planets on orbits exterior to a slow moving Jupiter will migrate rapidly inwards and may be captured into exterior mean-motion resonances (MMR) \citep{Lee:2002,Thommes:2005}.  By comparison, small interior planets may be shepherded into MMRs during the initial, fast migration phase of the Jupiter \citep{Zhou:2005,Raymond:2006}.  Thus, disk migration models often predict the presence of neighboring ``companion'' planets in or near MMR with a \hj.

These, small companions near interior or exterior MMRs would induce orbital perturbations that can be seen as transit timing variations (TTVs) about a constant period \citep{Agol:2005,Holman:2005}.  While tidal damping or other processes can displace the planets from resonance \citep{Terquem:2007}, near-resonant systems can still produce a large TTV signal \citep{Agol:2005,Holman:2005} and planets with masses much smaller than Jupiter may be detected through these variations.

Few companion planets are found in \hj\ systems---none in nearby orbits \citep{Wright:2009}.  Stability considerations may restrict orbits that are much closer than the 3:2 MMR.  Nevertheless, strong limits on resonant or near-resonant companions, with mass constraints smaller than the mass of the Earth near the 2:1 and 3:2 MMRs, exist from TTV studies \citep{Steffen:2005,Gibson:2009}, and nothing has turned up in searches for additional transiting companions to \hj s \citep{Croll:2007}.  Hot Jupiters are, however, known to have distant stellar or planetary companions \citep{Butler:1999,Eggenberger:2004}.  Yet, no evidence suggests that \hj s preferentially have companions capable of driving their orbits inward through Kozai cycles and tidal friction (contrary to predictions by \citep{Fabrycky:2007}), and the lack of near-resonant companions is at odds with nominal predictions of disk migration.  Still other arguments point out the importance of including interactions with distant planets \citep{Wu:2011,Naoz:2011}.  Thus, while some theories are fading into disfavor, the fundamental mechanism that produces the \hj\ population remains unexplained.


If hot Jupiters originate beyond $\gtrsim 1$ AU, somehow gain sufficient eccentricity to induce a tidal interaction with the star, and settle into their close orbits, then planets interior to 1 AU would be scattered during the gas giant's dynamical evolution.  Such a scenario would explain the lack of discoveries from TTV studies and photometric transit searches.  The latter issue was discussed by \citep{Latham:2011}.  We revisit that subject here and also conduct a basic TTV analysis on a large sample of \hj\ systems identified in the \kepler\ data in an effort to make definitive statements about the presence of nearby companions in a large sample of candidate systems.


\section{Sample Selection}\label{sample}


The main focus of this work is stars similar to the sun, we therefore exclude M dwarfs from our sample, which also have less reliable estimates of stellar properties.  The distribution of stellar temperatures of the \kepler\ Objects of Interest (KOIs) shows obvious bimodality 
since M dwarf stars were preferentially included in the target list for the mission.  We make a temperature cut at 4600K, only taking stars with temperatures, as reported in B11, greater than this value (this cut also excludes some late K-type stars).


We established selection criteria for planet sizes and periods in a similar fashion---identifying natural breaks in the distribution where a cut can be made\footnote{The results of the study depend very little on the precise location of the sample boundaries.}.  To choose the range of orbital periods we first select all KOIs that have sizes larger than 0.5 \rj\ and periods less than 30 days\footnote{The choice to use 0.5 \rj\ here was independent of the later adoption of 0.6 \rj\ for the lower boundary of the \hj\ sample.}.  The resulting distribution of orbital periods has a peak near 3.5 days and a noticeable trough just before 7 days.  
Using this information we choose planets with periods between 0.8 and 6.3 days.

\begin{figure}
\includegraphics[width=0.45\textwidth]{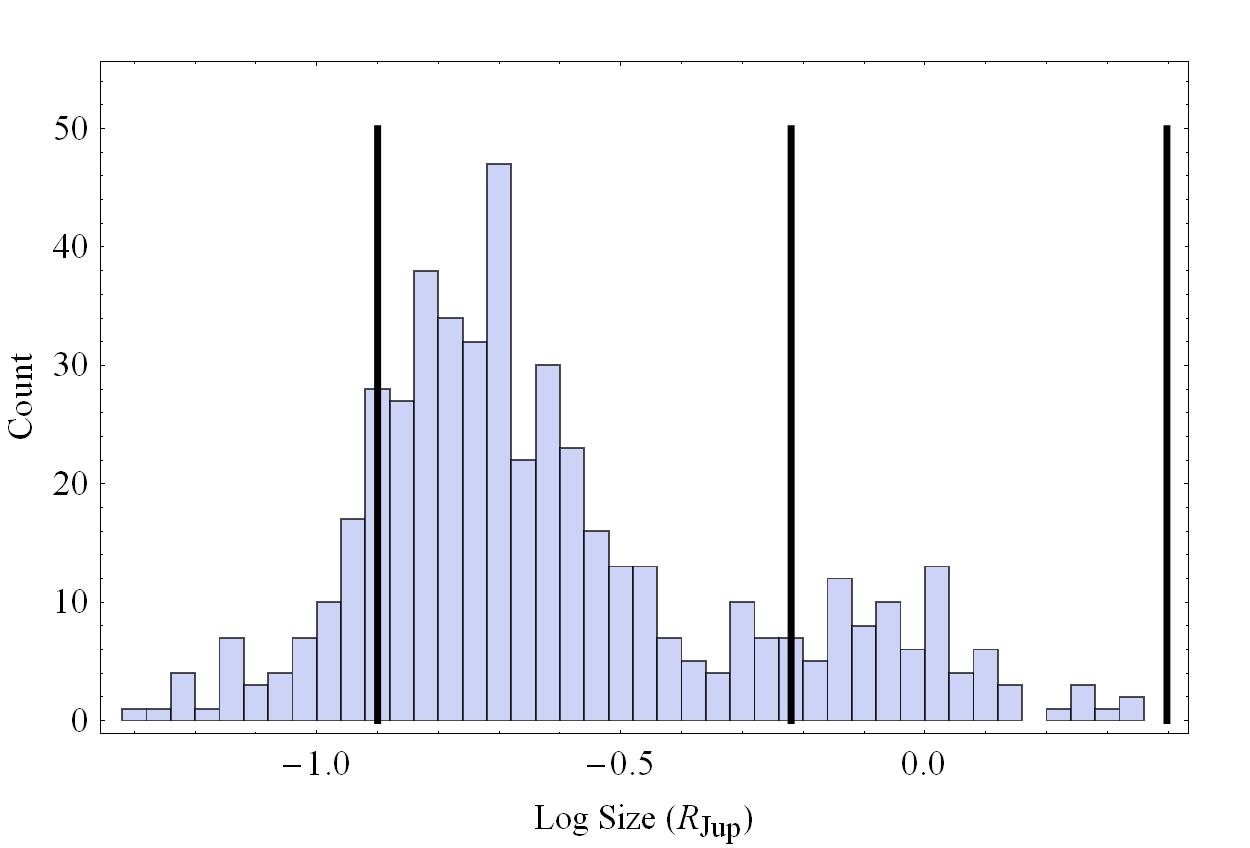}
\caption{Distribution of candidate planet sizes for all KOIs with periods between 0 and 10 days.  The rightmost partition, between 0.6 and 2.5 \rj\ is our size criterion for the \hj\ sample.  The middle partition, between 0.126 and 0.6 \rj , is used to select the hot Neptunes, and the leftmost partition, below 0.126 \rj , is used to select the hot Earth sample (defined in the discussion section).\label{sizecut}}
\end{figure}

We choose our boundaries for the planet sizes by first selecting all planet candidates with orbital periods between 1 and 10 days (see Figure \ref{sizecut}).  
We see a transition from Jupiter size objects to the much larger population of Neptune and smaller objects in the distribution of candidate sizes and choose \hj\ candidates with sizes between 0.6 and 2.5 \rj.  The number of KOIs that satisfy the above selection criteria is 63, and they constitute our \hj\ sample (we note that uncertainties in the stellar radii may produce systematic bias or uncertainty in these candidate sizes).

\begin{figure}
\includegraphics[width=0.45\textwidth]{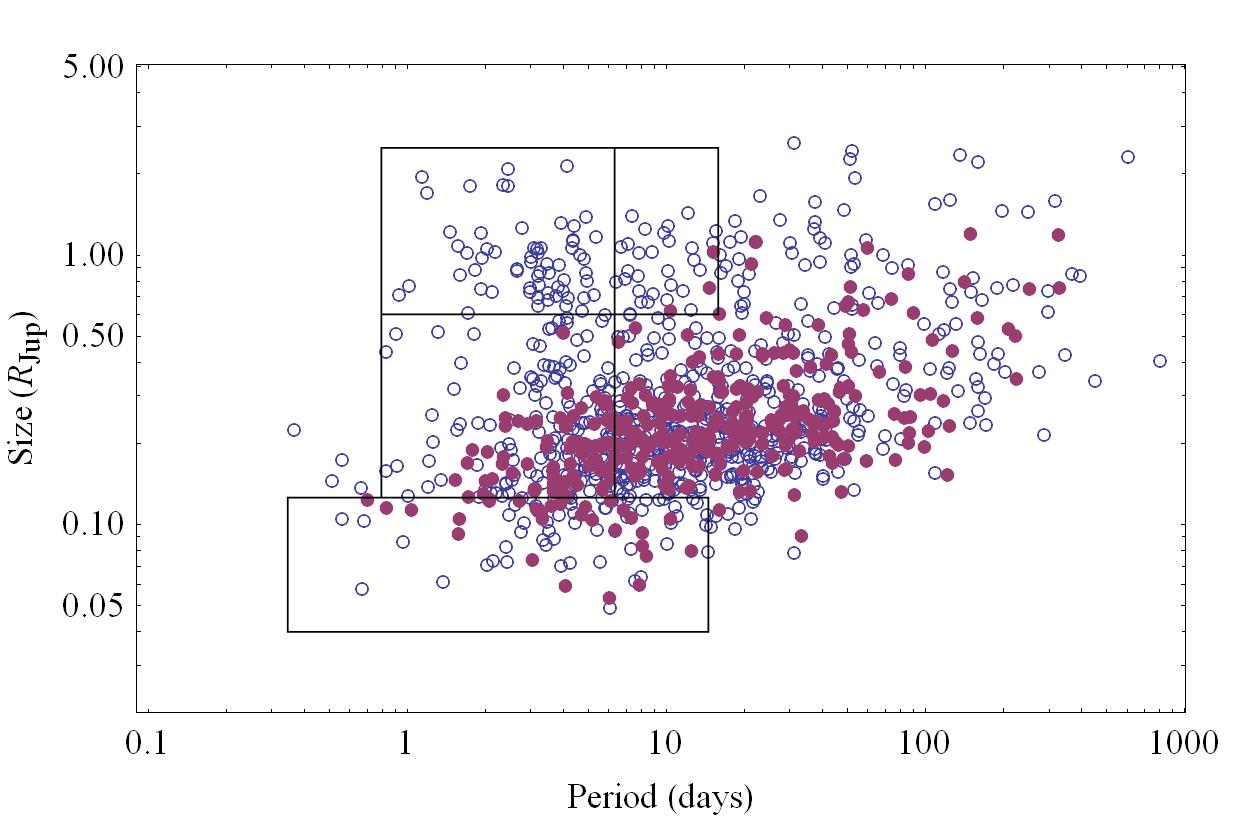}
\caption{Scatter plot showing the samples for \hj s (upper left box), warm Jupiters (upper right box), hot Neptunes (center box), and hot Earths (lower box).  KOIs in single transiting systems are the blue, open circles, while multiple transiting systems are red, filled circles.  The sizeable population of single, large planets stands out from the lack of red, filled circles in the upper, left portion of this plot.\label{fullsample}}
\end{figure}

In addition to the sample of \hj s, we consider two neighboring samples of KOIs, specifically hot Neptunes and warm Jupiters.  For the hot Neptunes, we select all KOIs with sizes between 0.126 and 0.6 \rj and periods between 0.8 and 6.3 days.  The warm Jupiters satisfy the same size criteria as the \hj s, but have periods between 6.3 and 15.8 days.  These cuts yield 224 hot Neptunes and 32 warm Jupiters.  In each of these samples there is one system that we ignore as they are missing several quarters of data.  Also, KOI-928.01, a known triple star system involving an eclipsing binary \citep{Steffen:2011} is excluded from the \hn\ sample.  This leaves 222 hot Neptune systems and 31 warm Jupiter systems.  Figure \ref{fullsample} is a scatter plot of candidate size vs. orbital period for KOIs given in B11 that are analyzed here, with the boundaries of the \hj\ and comparison samples shown.  There is a noticeable lack of planet candidates from multiple transiting systems for large planets on short orbital periods---where the \hj\ planets are defined.

\section{Companion search results}\label{searches}

For these samples, we look for evidence of additional companions whether by their transits or from dynamically induced TTVs.  These two searches can respectively constrain the sizes and masses of secondary planets in these systems.

\subsection{Transit search}

No additional planets have been found in any of the \hj\ systems.  However, using the combined differential photometric precision (CDPP) value for each system, we place an upper bound on the sizes of additional transiting planets that would be detected from the \kepler\ lightcurves.  CDPP is tabulated each quarter and effectively gives the mean photometric noise for that quarter in parts per million for a few specified durations (we use the 3-hour CDPP values here).  To estimate the size of planets that we are sensitive to for the different systems, we use the average of the CDPP values for quarters two through six for each target star.  The minimum detectable planet size is approximately given by
\begin{equation}
R_{\text{min}} = R_\star \left(\eta \frac{\text{CDPP}}{10^6}\right)^{1/2}\left(\frac{3}{ND}\right)^{1/4}
\label{CDPPequation}
\end{equation}
where $N$ is the number of transits, $D$ is the transit duration in hours, CDPP is for 3 hours in parts per million, and $\eta$ is the chosen detection threshold---we use 10 which formally gives $>99\%$ detection efficiency, though in practice it may be somewhat less ($\eta=7.1$ is the formal 50\% detection efficiency).

\begin{figure}
\includegraphics[width=0.45\textwidth]{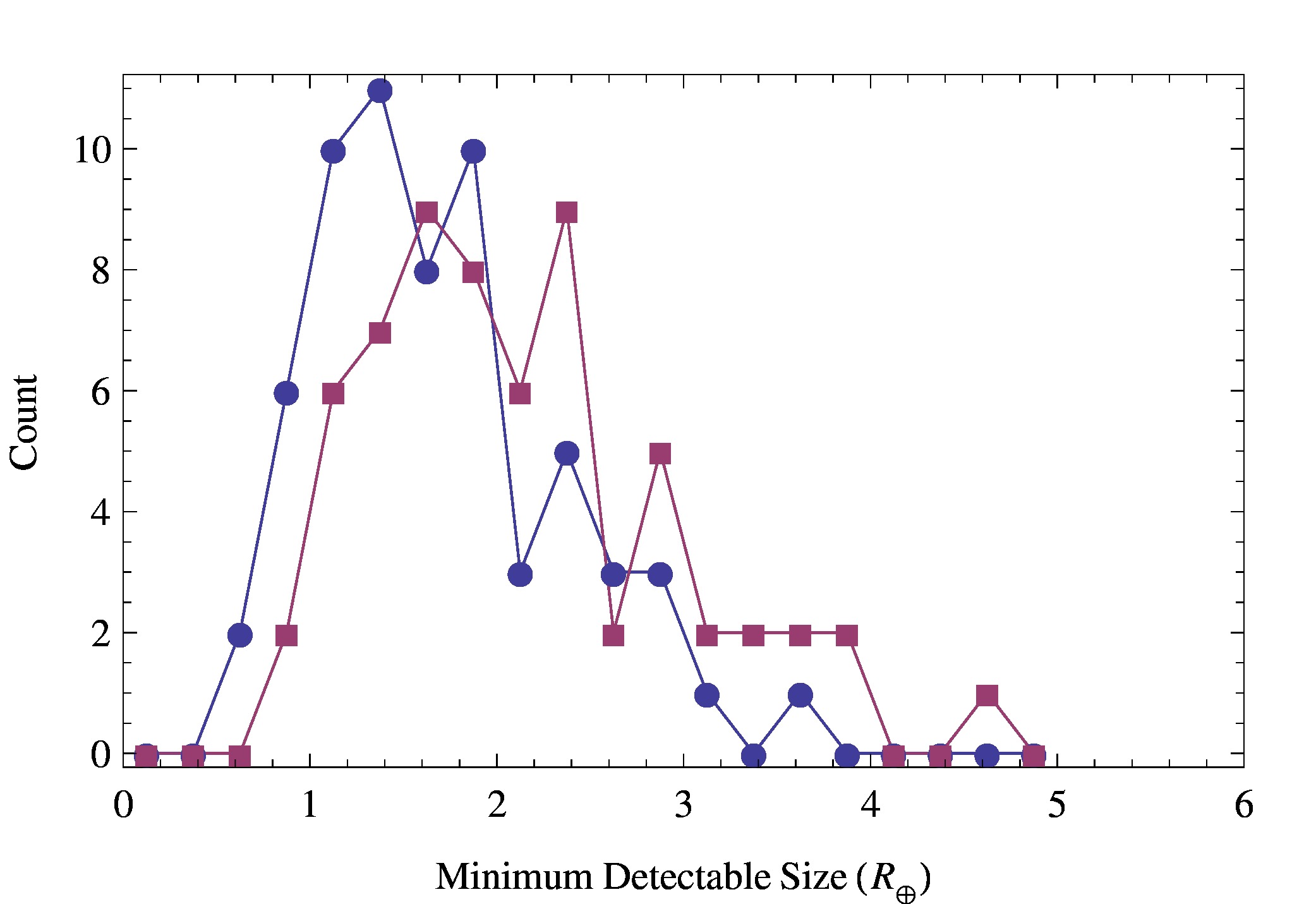}
\caption{The distribution of the minimum detectable planet size (in Earth radii) for the 63 KOIs in the \hj\ sample using Equation \ref{CDPPequation}.  The red (square) portion is for companions in the exterior 2:1 MMR and blue (circle) portion is for the interior 2:1 MMR.\label{sizes}}
\end{figure}

For the exterior 2:1 MMR, the largest and smallest detectable planets for all of the KOIs in the \hj\ sample are 4.7\re\ and 0.88\re\ respectively with a median of 2.0\re.  For planets with shorter orbital periods the size constraints become more stringent.  For example, the interior 2:1 MMR gives 0.70, 1.6, and 3.7 \re\ for the minumum, median, and maximum detectable planet sizes respectively.  Exterior planets would most likely come from their migration within the gas disk, while interior planets would come from shepherded planets by a migrating Jupiter.  The distribution in the minimum detectable sizes of planets in these systems are shown in Figure \ref{sizes} for the interior and exterior 2:1 MMR.

\subsection{Transit timing variations}

To search for TTV signatures in the \hj\ systems, we look for the best fitting sinusoidal model to the timing residuals after fitting for a constant period (the observed minus calculated, ``O-C'' residuals).  We then use an F-ratio test to determine whether the inclusion of the additional model parameters is justified given the data.  We note that a real TTV signal is the sum of several Fourier components each with its own amplitude and period.  
However, the largest TTV signals appear when the planets are near MMR and in those situations the signal is dominated by a single Fourier component.  

We measured the transit times following the analysis outlined in \citep{Ford:2011a} using the transit models from \citep{Mandel:2002}.  The current method of determining transit times occasionally results in outliers and points with unusually large error bars.  These discrepant points are generally caused by the presence of multiple, neighboring local minima in the transit fitting function.  Consequently, for each system we throw out any transit where the timing residual is larger than 5 times the median absolute deviation of all the timing residuals or where the error bars are five times the median of all error bars.  This conditioning typically eliminates few or no transit times.

We find evidence for significant TTV signals in two systems KOI-1177 and KOI-1382.  All others have a p-value for the F-ratio test greater than 0.1---indicating no compelling deviations from a constant period\footnote{We note that if we were attempting to claim the detection of a significant signal based upon this method, a monte carlo test of the significance of the measured p-value would be more appropriate.  The generic F-ratio test simply gives systems where further scrutiny is justified.}.  
We note that two systems in the \hj\ sample were identified in \citep{Ford:2011a} as potentially having TTVs in the first quarter of \kepler\ data.  KOI-10 had a slightly different linear ephemeris in early data from what was found through five quarters.  Additional data on KOI-10 did not continue that trend.  KOI-13 showed an early outlier transit time, which additional data confirms as an outlier.

Inspection of the lightcurves for KOI-1177 and KOI-1382 indicates that the observed TTVs in both systems are not due to planetary dynamics.  The residuals in KOI-1177 are primarily due to stellar variability causing the detrending algorithm to inject deviations in the measured times---application of a different detrending algorithm reduces the amplitude of the variation significantly.  The timing residuals in KOI-1382 have their peak power at the frequency equal to the difference between the observed star-spot modulation (or stellar rotation) frequency and the planet orbital frequency.  Thus, in both systems there is a natural explanation for the TTV signal that does not invoke an additional planet.





For those systems not showing TTVs, rather than giving specific calculations for the maximum allowed companion mass in each, we point out that numerical simulations show that an Earth-mass planet on a circular orbit near the 2:1 MMR can easily induce a TTV signal with $\sim 1$ minute amplitude on a Jupiter-mass planet with a 4-day orbit, and that in this regime the TTV signal scales linearly with the mass of the perturbing planet.  Thus, for these systems, where the timing uncertainty is between 0.1 and 15 minutes, the maximum allowed companion mass in or near a resonant orbit is between the masses of Mars and a few times the Earth.  Larger masses, two to three orders of magnitude larger, are allowed planets far from resonance.  However, such planets would typically have larger sizes and smaller orbital period variations---and would therefore likely be seen in the transit search described above unless there is a nearly universal tendency for large mutual inclination.

\section{Comparison with nearby populations}\label{compare}


\subsection{Warm Jupiters}

The \wj\ sample contains 31 objects and includes all KOIs with sizes between 0.6 and 2.5 \rj\ and periods between 6.3 and 15.8 days (see Figure \ref{fullsample}).  In this sample there are three objects that are known to be in multiple transiting systems, KOI-137.02 (Kepler-18d) \citep{Cochran:2011}, KOI-191.01 \citep{Steffen:2010}, and KOI-1241.02 whose companion is near the exterior 2:1 MMR.  All three of these objects are near the long-period periphery of the selection region.

The TTV analysis of this sample produces three systems with plausibly significant TTV signals (meaning the p-value of the F-ratio test is less than 0.1): Kepler-18d, 190.01, and 1003.01.  Kepler-18d was known to have a large TTV signal due to its Neptune-size companion near the interior 2:1 MMR (this companion, Kepler-18c, lies just outside of our allowed periods for the \hn\ sample).  KOI-190.01 has a TTV signal that is quite similar to what is observed in Kepler-18d and may have an unseen perturbing companion.

The fact that that at least 5 of the 31 warm Jupiter systems show some evidence of a companion implies that $\gtrsim 10\%$ of \wj\ systems have such companions.  These additional companions can be seen either from their transits, from their dynamical influence as in KOI-190 (which has no known transiting companion), or both as in Kepler~18.  This draws a sharp contrast with the \hj\ candidates which have similar sizes, slightly shorter orbital periods, and no evidence for companions even with a sample that is twice as large.


\subsection{Hot Neptunes}

The \hn\ sample contains the 222 KOIs with periods between 0.8 and 6.3 days and sizes between 0.126 and 0.6 \rj (see Figure \ref{fullsample}).  In the sample of \hn s there are 73 (roughly 1/3 of the sample) that are known to have additional transiting objects.  The TTV analysis shows two systems with significant TTV signals: KOI-244.02 
and KOI-524.01 (which has no visible compaion).  Taking all of the systems in this sample, there are 84 companion planets whose orbital periods are within a factor of 5 of the \hn\ that marked their selection\footnote{Several in this sample have multiple companions in closely packed systems.  So there are more pairs than there are sample members.} and 38 with period ratios within a factor of 2.3 (the choice of 2.3 is explained in the discussion).  These observations further indicate that the \hn\ systems are quite different from \hj\ systems (as noted in RV studies by \citep{Mayor:2008}) where a large fraction of systems have multiple planets and that planet pairs are often in close proximity.  While most of these companions are known from their transits, some have been detected solely from their TTV signal.  The fact that a smaller fraction of the \hn\ systems shows TTVs than the \wj\ systems is due in part to the worse timing precision and smaller TTV signal of the smaller and less massive planets.








\section{Discussion}\label{discussion}

There are a few possible explanations for the lack of observed companions in \hj\ systems: 1) they might not exist; 2) they may exist, but are yet too small to have been seen; 3) they may exist, but have very large TTVs and are missed by the transit search algorithm (which assumes a nearly constant orbital period); and 4) they may exist, but have been scattered into highly inclined orbits, and therefore are unlikely to transit.


\subsection{No companions}

The first reason why companions to \hj s are not observed is that they simply may not exist in large quantities at the present time.  Such small planets may have formed in the systems and been subsequently ejected through planet-planet scattering, pushed into the star through a combination of shepherded migration and tidal dissipation of orbital energy (via the induced eccentricity from the giant), or by some other means.  Another option is that \hj\ systems form differently than the majority of planetary systems such that small planets are simply not produced.

\subsection{Small sizes}

A second possibility for lack of small companions is that companions that survive today are below the detection threshold of the \kepler\ spacecraft and the current transit search pipeline.  The results of our CDPP analysis above show that two of the \hj\ systems show no planets larger than the Earth and more than half (32 of the 63) show no companions larger than twice the Earth for any orbital period out to the exterior 2:1 MMR with the \hj.

If small, but detectable planets exist in some systems, then we can estimate a reasonable maximum for the fraction of systems that have them.  Suppose some fraction of \hj\ systems do have nearby companions and that we were unlucky that no examples appear in our sample.  The Poisson probability of zero events occurring is 0.05 for a distribution with a mean of 3.  This implies, with 95\% confidence, that no more than 3 of 63 \hj\ systems (or 5\%) can have nearby detectable companions.  Ultimately, more data will allow us to constrain the presence of companions with smaller sizes.

\subsection{Small masses}

Since no obvious TTVs are visible in the \hj\ systems, it is necessary that any perturbing planets have small masses or are in orbits where the TTV signal is much smaller over the timescale of these data.  Since the observed objects are Jupiter size, the timing precision of their transits is quite good, the median being 70 seconds.  Existing analyses of TTV signals with slightly worse timing precision and far fewer transits (e.g., \citep{Steffen:2005} had 100-second timing precision and 11 transits), have sensitivity to masses smaller than the Earth.  \kepler 's improved timing precision and temporal coverage allows for the detection of planets approaching that of Mars (see \citep{Agol:2007}).  A rocky object with a mass this small may not appear in the photometry through Q6.


Initially one would expect shepherded objects to be near MMR, but perturbations to the orbit from the \hj\ combined with tidal dissipation may cause a drift from resonance.  If the perturber were far from resonance, then photometric constraints are more powerful than TTV constraints as the mass sensitivity of TTVs to such objects can fall by two to three orders of magnitude---closer to the mass of Neptune ($\gtrsim 20$ \me).  However, only unphysically dense planets can have masses that large and yet be undetected in transit.

Should low mass companions be missed by the transit detection software due to their own TTVs, a transit search method that allows for a varying period could be employed to identify them.  However, since the number of expected transits for planets with such small orbital periods is quite large, very few objects of sufficient size could escape detection by the existing transit identification pipeline since even with variations in the orbital period, several of the transits would still be well fit by a constant-period model.

\subsection{Large mutual inclinations}

Another explanation for the lack of companions is that orbits in these systems might have large mutual inclinations.  Rossiter-McLaughlin measurements of the obliquity of \hj\ planetary orbits (the angle between the planet orbital axis and the stellar rotation axis) show that highly misaligned configurations are not rare \citep{Triaud:2010,Winn:2010,Morton:2011}.  It is reasonable to expect small companions might exist in highly inclined orbits with respect to the orbital plane of the transiting candidate.

\begin{figure}
\includegraphics[width=0.45\textwidth]{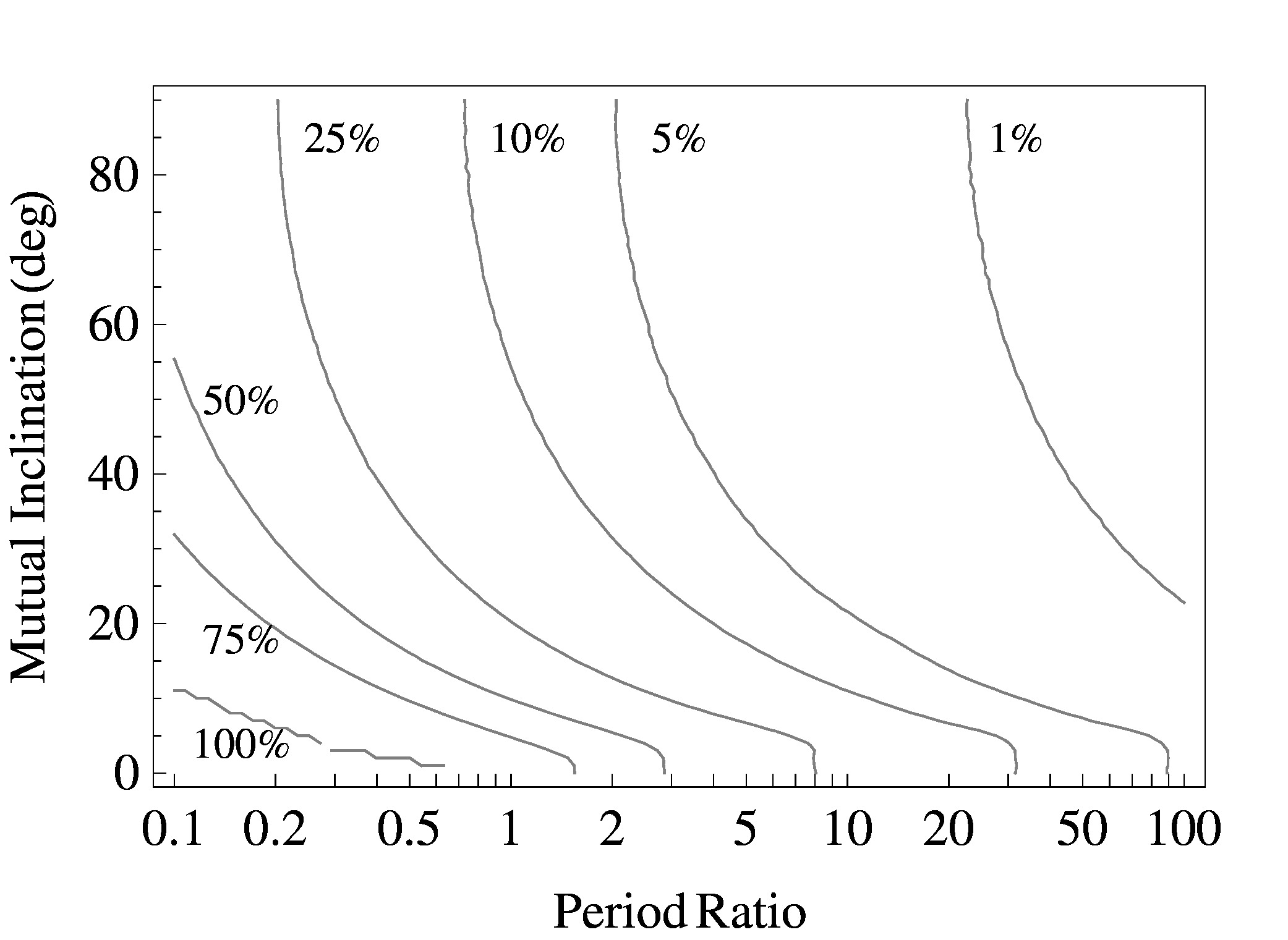}
\caption{The percentage of the hot Jupiter sample that would show a second planet transiting as a function of period ratio and mutual inclination from Monte Carlo simulation.  For example, if every hot Jupiter had a detectable companion near the exterior 2:1 resonance with a mutual inclination of 40 degrees, the expected number of hot Jupiters with transiting companions would be 8\% or 5 out of 63.
\label{transprob}}
\end{figure}

Suppose all hot Jupiters have a detectably-large companion whose orbit has a large mutual inclination.  A randomly placed observer would either see neither, either, or both planets (if looking down the line of nodes \citep{Ragozzine:2010}).  To quantify the latter case, Figure \ref{transprob} shows a Monte Carlo simulation of the geometric probability that a companion to a \hj\ would transit as a function of period ratio and mutual inclination.  
Even if the companion was on a perpendicular orbit, random viewing orientations would yield transits of the companion in $\sim$13\% of systems ($\sim 8$ detections) at the interior 2:1 MMR and 5\% ($\sim 3$ detections) at the exterior 2:1 MMR.  Thus, high mutual inclinations cannot entirely explain the lack of observed companions---they must either be infrequent or too small.

Even should only a portion of the \hj\ systems have highly inclined companions, we can still constrain that fraction.  Using Poisson statistics, at the interior 2:1 MMR not more than $\sim$40\% of \hj\ systems---a fraction similar to the fraction of observed companions in the \hn\ sample---can have a companion on a perpendicular orbit (at the 95\% confidence level).  No more than 60\% of hot Jupiters can have detectable planets on interior orbits at any mutual inclination, with much more stringent constraints ($\lesssim 5$\% can have such companions) for mutual inclinations similar to the \hn\ systems of a few degrees \citep{Lissauer:2011b}, casting serious doubt on models that predict such planets (e.g, \citep{Zhou:2005}).

Another way to directly test for large mutual inclinations is to look for TTVs in systems with single hot Earths (Earths and super Earths).  The TTVs would be induced by the presumed presence of a non-transiting \hj\ companion---and would be much larger than the TTVs induced on the \hj\ by the smaller object.  We selected a ``hot Earth'' sample from the planet candidates comprising all KOIs in single systems with radii less than 0.126 \rj\ (1.4 R$_\oplus$) and orbital periods between 0.34 and 14.5 days (a factor of 2.3 smaller and larger than the \hj s since TTV signals are largest within these period ratios).  There are 53 such systems, though one system has significant gaps in the coverage---leaving 52 for study.  We note that 29 hot Earths in multiple systems satisfy this period criterion (over 1/3 of the total, similar to the \hn\ sample).

\begin{figure}
\includegraphics[width=0.45\textwidth]{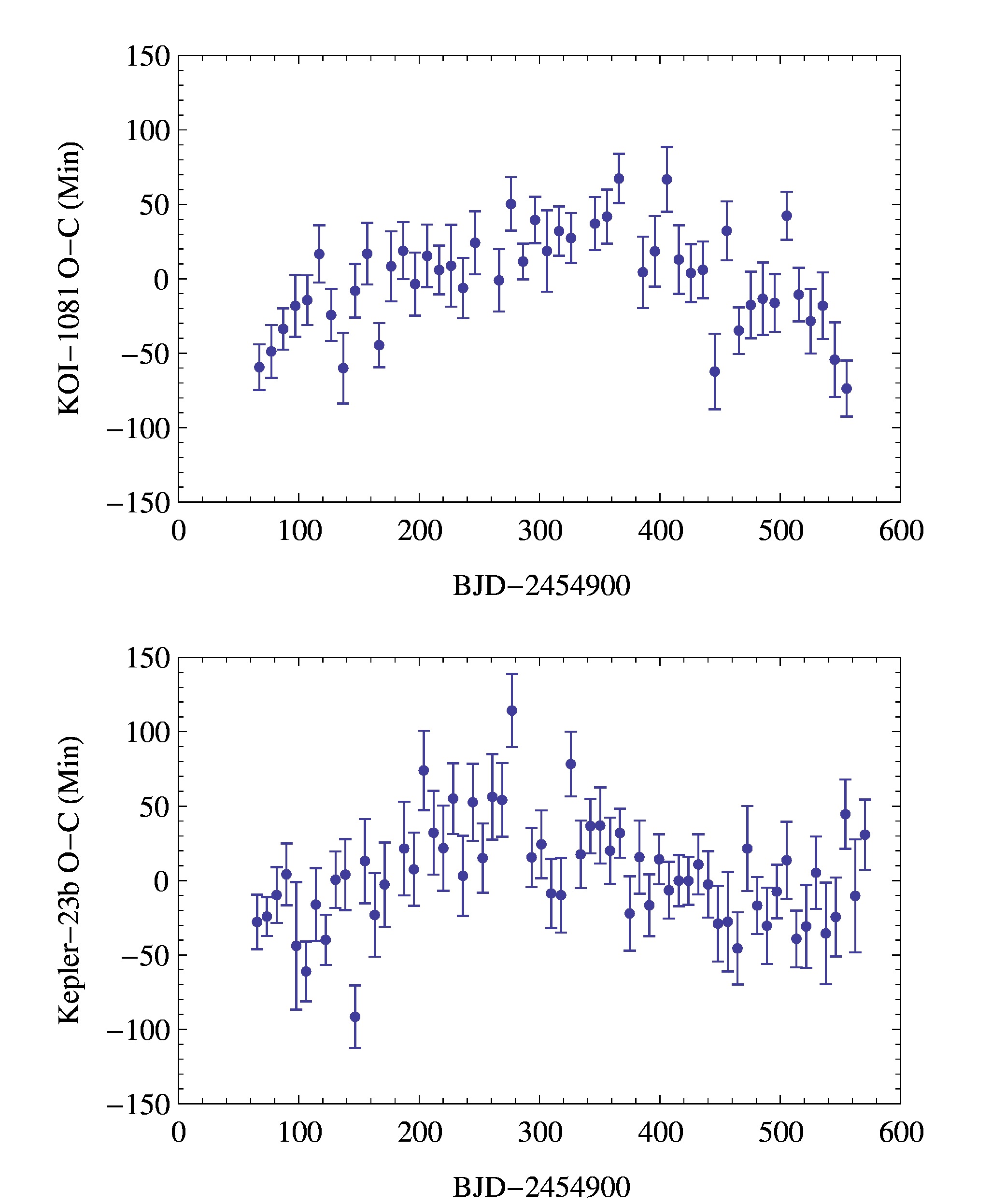}
\caption{The timing residuals from a constant period for KOI-1081.01 (top) and Kepler-23b (bottom).  KOI-1081 is the only system in the sample of isolated hot Earths where the F-ratio test rejects a constant period model ($p=0.013 \ (< 0.1)$) while Kepler-23b is an example of a hot Earth in a known multiple system with a similar TTV signal induced by a small, near-resonant planetary companion. \label{koi1081}}
\end{figure}

We searched for significant TTV signatures in the single hot Earth systems, finding one system where the $p$ value of the F-ratio test is less than 0.1 (KOI-1081 with $p=0.013$).  A plot of the TTV signal for KOI-1081, which has an estimated size of 0.125 \rj\ and a period of nearly 10 days, is shown in Figure \ref{koi1081}.  We do not attempt a detailed analysis of this TTV signal here, only pointing out that it exists and may be due to an unseen, interior, Jupiter-size companion.  However, we note that a similar analysis of the 29 hot Earths in multiple systems shows one object, KOI-1102.02 (Kepler-23b \cite{Ford:2012}), with a similar orbital period (8.1 days), a similar $p$ value (0.028), and a TTV signal with similar amplitude and duration that is caused by its small known transiting companion near the 3:2 MMR (also shown in Figure \ref{koi1081}).  These similarities suggests that the observed TTV signal in the isolated KOI-1081 system might be due to a nontransiting, near resonant planet with smaller size, as is the case with Kepler-23.

The typical timing error for the sample of hot Earths is about a factor of 20 larger than for the \hj's (the median being 0.02 days or 30 minutes).  Consequently, the sensitivity to companion mass is much worse.  However, as we are testing for the presence of a non-transiting \hj\ and the expected mass of the pertuber, and its associated TTV signature, is of order 100 times larger, the lack of observed TTVs in this Earth sample is a particularly stringent constraint on the presence of hot Earth/\hj\ systems.  We note that since an exhaustive study of the TTV signal with mutually inclined orbits does not appear in the literature, there may be some configurations where the orbital elements of the system conspire to hide the TTV signal.  Such singular configurations are, of necessity, quite rare.  If many systems are in those configurations then some dynamical mechanism would be required to drive the systems into those exotic orbits.

\section{Conclusions}\label{conclusions}

Neither a photometric search nor a TTV search yields compelling evidence for nearby companion planets to \hj s (within a factor of a few in orbital period), in any of our sample of 63 candidate \hj\ systems.  While such planets may yet exist, they must be either very small in size ($\lesssim 1 R_\oplus$) or mass ($\lesssim 1 M_\oplus$ for near resonant planets).  Nonresonant planets with small masses or sizes are still allowed, as are planets with much longer orbital periods.  A TTV study of hot Earths shows no significant evidence for high-mass companions on inclined orbits---effectively eliminating mutually inclined orbits as the reason for the lack of detected companions.  Here again, however, planets with small masses and small sizes are allowed.

Both the photometric search and the TTV search for companions in neighboring size and period bins turn up positive results.  Roughly 1/3 of the 222 \hn\ systems are in multi-transiting systems and two show significant TTV signals.  Three of 31 warm Jupiter systems have transiting companions.  Two of these three show TTVs along with one system without a known transiting companion.

The presence or lack of companions in \hj\ systems is a distinguishing characteristic of planet formation and dynamical evolution theories.  The definitive lack of neighboring Earth-size and Earth-mass companions in \hj\ systems favors formation models of involving eccentricity excitation followed by tidal circularization \cite{Rasio:1996,Weidenschilling:1996,Holman:1997,Wu:2003,Fabrycky:2007,Nagasawa:2008,Wu:2011,Naoz:2011}.  The presence of additional companions to \hn s and hot Earths suggests that most short-period, low mass planets have a different formation history from \hj s.  Moreover, the combination of few companions to \hj s and frequent companions to low-mass short-period planets indicates a mass dependence in system architecture.  This dependence on planet mass suggests \hj\ formation often occurs from planet-planet scattering because eccentricity excitation by planet-planet scattering is mass dependent while excitation by a wide binary companion is not.

Hot Jupiter systems where planet-planet scattering is important are unlikely to form or maintain terrestrial planets interior to or within the habitable zone of their parent star.  Thus, theories that predict the formation or existence of such planets \citep{Raymond:2006,Mandell:2007} can only apply to a small fraction of systems.  Future population studies of planet candidates, such as this, that are enabled by the \kepler\ mission will yield valuable refinements to planet formation theories---giving important insights into the range of probable contemporary planetary system architectures and the possible existence of habitable planets within them.





\begin{acknowledgments}
Funding for the \kepler\ mission is provided by NASA's Science Mission Directorate.  We thank the \kepler\ team for their many years of hard work.  J.H.S acknowledges support from NASA under grant NNX08AR04G under the Kepler Participating Scientist Program.  D. C. F. and J. A. C. acknowledge support from NASA through Hubble Fellowship grants \#HF-51272.01-A and \#HF-51267.01-A awarded by the Space Telescope Science Institute, operated by the Association of Universities for Research in Astronomy, Inc., under contract NAS 5-26555.
\end{acknowledgments}





\bibliography{multis}

\end{document}